\documentclass[sigconf, review]{acmart}

\AtBeginDocument{%
  }

\makeatletter
\renewcommand\@formatdoi[1]{}
\makeatother

\acmConference[]{}{}{}

\renewcommand\footnotetextcopyrightpermission[1]{} 
\makeatletter
\@ACM@reviewfalse
\@ACM@anonymousfalse
\makeatother

\begin{document}

\title{Non-Invasive Arterial Pulse Detection with Millimeter-wave Radar and Comparison With Photoplethysmography}

\author{Nima Bahmani}
\affiliation{%
  \institution{Aalto University}
  \city{Espoo}
  \country{Finland}}
\email{nima.bahmani@aalto.fi}

\author{Dariush Salami}
\affiliation{%
  \institution{Aalto University}
  \city{Espoo}
  \country{Finland}}
\email{dariush.salami@aalto.fi}

\author{Hüseyin Yiğitler}
\affiliation{%
  \institution{Aalto University}
  \city{Espoo}
  \country{Finland}}
\email{yusein.ali@aalto.fi}

\author{Juhapekka Hietala}
\affiliation{%
  \institution{Aalto University}
  \city{Espoo}
  \country{Finland}}
\email{juhapekka.hietala@aalto.fi}

\author{Tuukka Panula}
\affiliation{%
  \institution{University of Turku}
  \city{Turku}
  \country{Finland}}
\email{tuukka.j.panula@utu.fi}

\author{Stephan Sigg}
\affiliation{%
  \institution{Aalto University}
  \city{Espoo}
  \country{Finland}}
\email{stephan.sigg@aalto.fi}

\renewcommand{\shortauthors}{Bahmani et al.}

\begin{abstract}
Cardiovascular diseases remain a leading cause of mortality and disability. The convenient measurement of cardiovascular health using smart systems is therefore a key enabler to foster accurate and early detection and diagnosis of cardiovascular diseases and it require accessing a correct pulse morphology similar to arterial pressure wave. This paper investigates the comparison between different sensor modalities, such as mmWave and photoplethysmography from the same physiological site and reference continuous non-invasive blood pressure devide. We have developed a hardware prototype and established an experiment consist of 23 test participants. Both mmWave and PPG are capable of detecting inter-beat intervals. mmWave is providing more accurate arterial pulse waveform than green photoplethysmography.
\end{abstract}

\keywords{Wearable computing, Mobile smart systems, Arterial stiffness, mm-Wave, PPG, Arterial pulse wave}


\maketitle

\section{Introduction}

Blood pressure, a fundamental physiological parameter, serves as a crucial indicator of the cardiovascular system's function and is a key vital sign monitored in clinical settings~\cite{RN19,RN120}.
Its accurate and continuous assessment is important for the early detection, diagnosis, and management of cardiovascular diseases, which is still one of the most important cause of disability~\cite{RN74,RN92,RN119}.
Traditional methods of blood pressure monitoring, often relying on intermittent cuff-based measurements, have limitations to provide continuous blood pressure fluctuations and can be inconvenient for routine or overnight monitoring~\cite{RN19,RN92,RN120,RN40,RN41}.

Arterial stiffness, a measure of the elasticity of arterial walls, is increasingly recognized as a significant factor in cardiovascular health~\cite{RN95,RN141,RN89,RN41}.
The morphology of the arterial pulse wave is recognized as a fundamental component of a comprehensive evaluation of vascular health, offering additional health information. 
Changes in arterial stiffness are associated with aging, hypertension, and various other cardiovascular disorders, influencing the morphology of the arterial pulse wave~\cite{RN95,RN92,RN119}.
This complex waveform contains essential information about the physical properties of the arterial tree, including vascular stiffness~\cite{chen1997estimation}. 
Furthermore, changes in arterial stiffness lead to increased pulse wave velocity and the earlier return of reflected waves, which significantly alter the arterial pulse wave morphology~\cite{nichols2005clinical,nichols2008effects}. 
These structural and functional alterations in the pulse wave contour impose an increased workload on the cardiovascular system, thereby emphasizing why accurate characterization of arterial pulse wave morphology is necessary for a robust cardiovascular evaluation~\cite{nichols2008effects}.

In response to this need, wearable devices for continuous and non-invasive monitoring of physiological parameters, including cardiovascular signals, have emerged. Notably, wearable millimeter-wave (mmWave) radar and photoplethysmography (PPG) have gained attention for their potential in continuously capturing detailed cardiovascular information.
The correct comparison of these two sensor modalities can demonstrate the potential and limitations of these  methods.

We propose to integrate mobile PPG and mmWave sensing to monitor arterial pulse wave from the same measurement site of the human body.
The integration of wearable solutions that combine mmWave radar and PPG can leverage the unique strengths of both technologies, potentially leading to a more comprehensive and robust analysis of arterial pulse waves and making comparison between different sensor modalities. By enabling the continuous capture of detailed arterial pulse waveforms and related physiological data, these wearable technologies hold significant promise for the earlier detection and more effective management of prevalent conditions such as hypertension and other cardiovascular disorders~\cite{RN55,RN56,RN119}.
Our contributions are three-fold:
\begin{enumerate}
 \item we have developed a hardware prototype for mobile cardiovasular health monitoring, which integrates reflective PPG and mmWave sensing elements as interdisciplinary approaches;
 \item we report from a study with a diverse distribution of 23 healthy subjects. Particularly, we analyse the dynamic signals from both sensor modalities and evaluate the capability of each sensor solution to track subtle changes in the arterial pulse waveform which enables comparing the capability of mmWave and PPG in tracking the arterial pulse wave;
 \item we investigate the accuracy of the inter-beat interval (IBI) estimation as one example application.
\end{enumerate}

\section{Related Work}
Wearable PPG is a widely used optical technique for continuous heart rate monitoring and the assessment of blood volume changes. 
Niwarthana et al.~\cite{Niwarthana} proposed a novel deep learning approach utilizing mathematically modelled PPG waveform characteristics to achieve efficient and robust heart rate estimation from noisy wearable PPG sensors.
Neigel et al.~\cite{Neigel} demonstrated the feasibility of using consumer wearable PPG sensor to identify periods of stress within university students by observing changes in physiological biomarkers like heart rate and heart rate variability correlated with academic and societal events in a real-world environment.
Ananthan et al.~\cite{Ananthan} investigated the variations in signal characteristics such as IBI among students across different academic courses using wearable sensor data, revealing distinct patterns associated with course type and individual differences.

Recent advancements in near-field radio frequency (RF) sensing have demonstrated the potential for non-invasive biomedical measurements. Wearable mmWave radar offers the distinct advantage of being capable of detecting subtle tissue deformation resulting from underlying arterial pulsations. A notable area of focus involves the use of mmWave radar for the detection of arterial pulse waveforms.
Johnson et al.~\cite{RN37,RN55} presented a wearable mmWave device designed for the measurement of arterial pulses on the wrist for pulse wave analysis (PWA) and pulse transit time (PTT) in the context of blood pressure monitoring.
The mmWave radar was simulated and the tissue deformation was considered as the main stimulus and high level of agreement was reported between the mmWave and a tonometer. Wen et al.~\cite{RN19,RN15} also proposed a noninvasive and continuous blood pressure estimation developed by a 120 GHz compact antenna radar, extracting the radar-based pulse transit time (RPTT) for blood pressure prediction.

Beyond the wrist, near-field RF sensing is versatile to capture cardiovascular signals from different anatomical sites. The high frequency of mmWave signals also enables the detection of minute physiological movements with high sensitivity. Das et al.~\cite{RN117} presented a feasibility study on the non-invasive and continuous measurement of the jugular venous pulse using Frequency Modulated Continuous Wave (FMCW) radar. Li et al.~\cite{RN197} presented a near-field cardiac signal monitor based on radar which can be used as a chest wearing device.

Continuous non-invasive blood pressure (BP) measurement using RF techniques, particularly mmWave radar, has seen a considerable progress. Vysotskaya et al.~\cite{RN92,RN143} have demonstrated transformer-based approaches for accurate continuous BP estimation utilizing a 60 GHz radar.
Shi et al.~\cite{RN120} introduced a contact-free mmWave radar-based system for BP measurement with a focus on motion robustness. Tseng et al.~\cite{tseng2020noncontact} explored cuff-less blood pressure measurement using a microwave near-field self-injection-locked wrist pulse sensor. Further research has delved into specific radar technologies and their application in cardiovascular monitoring.
Lauteslager et al.~\cite{RN24} explored coherent UWB radar-on-chip systems for in-body measurement of cardiovascular dynamics.

Radar signal processing and algorythm plays a crucial role in extracting physiological information from radar signals. Shi et al.~\cite{RN120} proposed a delay-Doppler domain feature transformation method and a temporal referential functional link adaptive filter to enhance signal quality and mitigate the impact of motion on pulse waveform construction in their mmBP system. Vysotskaya et al.~\cite{RN92} also proposed a beamforming algorithm to improve the signal-to-noise ratio in radar-based blood pressure estimation. In addition to pulse wave measurements for blood pressure, near-field RF sensing has been applied to other vital signs, including cardiac and respiratory waveforms, systemic pulses, muscle contraction, eye movement, and tissue vibration~\cite{RN145}.

The simultaneous use of RF and PPG can provide complementary information regarding cardiovascular dynamics. PPG and radar sensors can be attached to separate sites on the body. For instance, Ebrahim et al.~\cite{pour2019blood} estimated blood pressure using on-body continuous wave radar in conjunction with PPG under various posture and exercise conditions. The continuous wave radar was implemented on the chest and the PPG was attached into the earlobe of the test participants. Gomes et al.~\cite{RN95}, have compared wrist-worn mmWave radar with fingertip PPG and a continuous non-invasive blood pressure (CNIBP) reference device. Hellbrück et al.~\cite{RN80} conducted brachialis pulse wave measurements using ultra-wide band (UWB) and continuous wave (CW) radar, alongside PPG and ultrasonic Doppler sensors. 

Building upon the established utility of combining RF/radar and PPG modalities for cardiovascular monitoring, often employing sensors at disparate anatomical sites, a gap persists concerning a direct, quantitative comparison of arterial pulse waveform morphology captured simultaneously by reflective PPG and near-field mmWave sensing when captured side-by-side from the identical anatomical location.

\section{Materials and Methods}



\begin{figure}
\centering
\resizebox{1\linewidth}{!}{\includegraphics{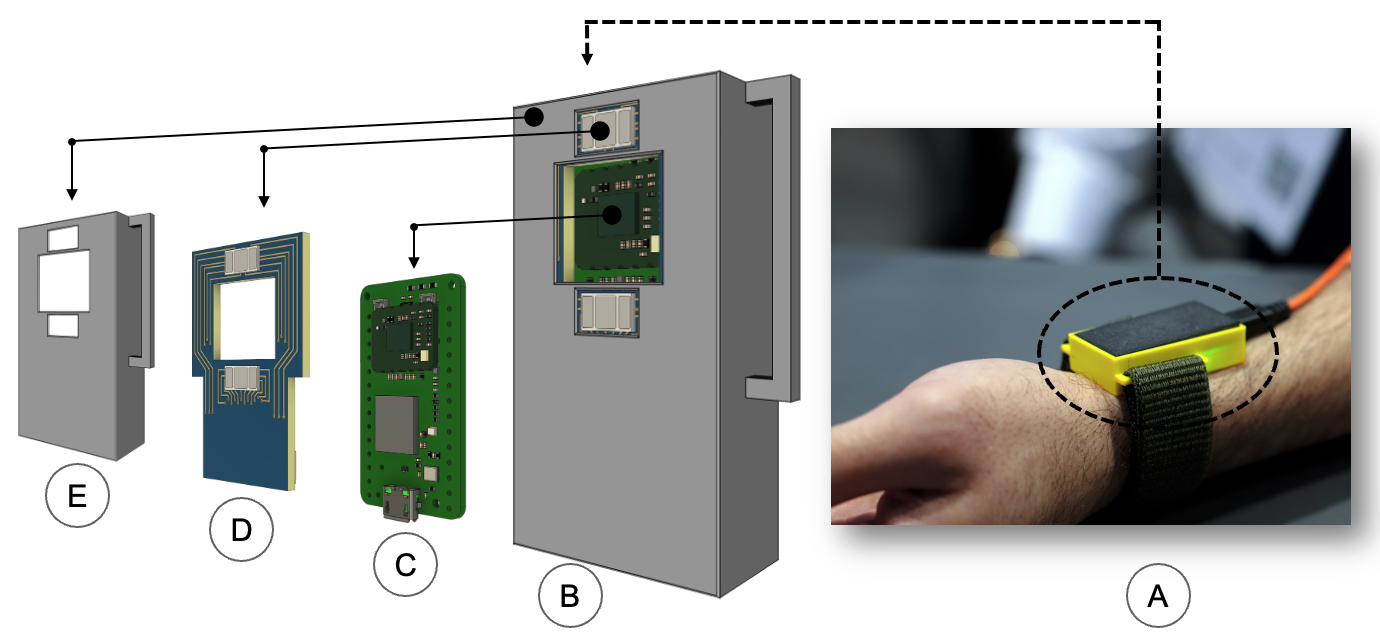}}
\caption{Prototype of the multimodal arterial pulse wave sensing device. (A) The wearable sensor positioned over the radial artery at the wrist. (B) The assembled prototype enclosure housing all sensor components. (C) mmWave radar module for near-field pulse wave sensing, (D) PPG optical front-end, and (E) the protective enclosure featuring a dedicated slot for wrist strap attachment.}
\label{fig_prototype}
\end{figure}

\begin{figure}
\centering
\includegraphics[width=0.9\linewidth]{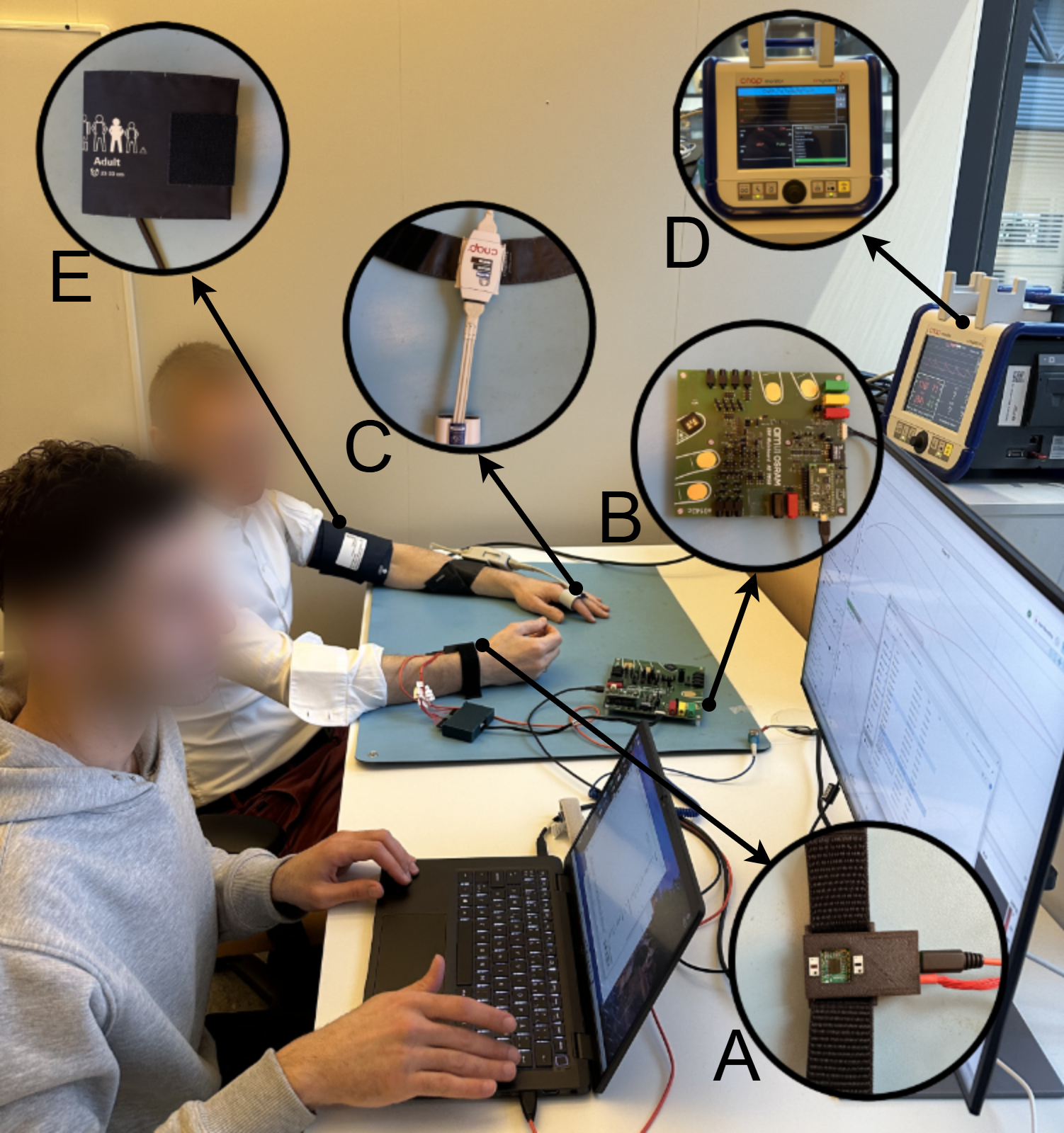} 
\caption{Experiment setup: (A) Multimodal sensing prototype, (B) ams-OSRAM 7058, (C) CNAP finger cuff, (D) CNAP 500, and (E) Arm cuff.}
\label{fig:sub1}
\end{figure}

The multi-modal prototype was developed in a wristband form factor for data collection, which is illustrated in Fig.~\ref{fig_prototype}. The mmWave sensing component of our prototype is based on the Infineon-BGT60TR13C evaluation kit.
This radar operates at 60 GHz with more than 5 GHz bandwidth and integrates three receive (RX) and one transmit (TX) antenna.
The BGT60TR13C evaluation kit is highly configurable, and its accompanying Software Development Kit (SDK) enables real-time data analysis.
We set the frame rate of the radar to 200~Hz to ensure that fine details within the pulse waveform are adequately sampled and captured.
For the optical sensing modality, our prototype incorporates the ams-osram AS7058 analog front end and its extension board.
The AS7058 is capable of driving multiple Light Emitting Diodes (LEDs) and reading signals from multiple photodiodes.
The extension board connects the optical front end to the AS7058.
We utilize SFH 7072 optical modules, which integrate LEDs of different wavelengths and photodiodes.
In the study we primarily utilize the green wavelength from one of the optical modules.
A dedicated Printed Circuit Board (PCB) was designed to solder the optical module and route the necessary electrical traces to connect to the AS7058 extension board.
The PCB design also includes an opening that strategically positions the optical modules for direct contact with the tissue, while ensuring that there is no barrier between the radar antenna and the tissue (see figure~\ref{fig:sub1} (A)).
The enclosure of our prototype positions the mmWave antenna at the desired 3mm distance from the tissue while attaching the optical modules to the skin surface.

As ground truth for validation, we utilize the industry gold standard, CNSystems Continuous Non-invasive Arterial Pressure (CNAP) consisting of a main display unit, a brachial cuff and a wrist-worn finger cuff unit.
The pressure values of the finger cuff are calibrated to brachial BP by the system firmware after an arm cuff measurement is performed.

This device provides a continuous measurement of the arterial pressure waveform, which is essential for evaluating the accuracy and fidelity of the signals captured by our integrated multimodal sensor.

\subsection{Participants}
\begin{figure}
\centering
\resizebox{1\linewidth}{!}{\includegraphics{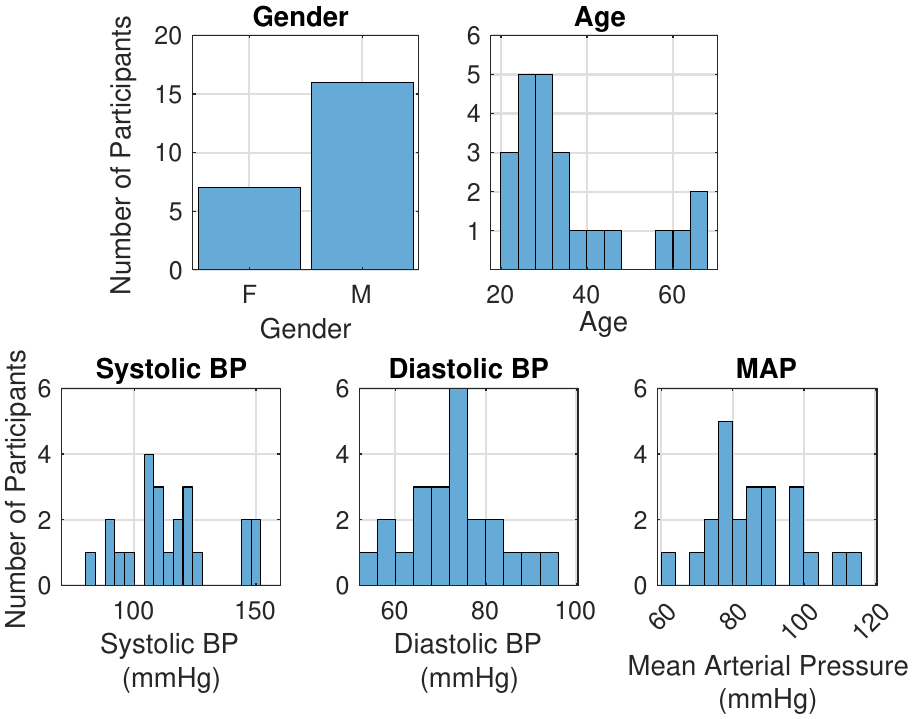}}
\caption{Participant demographics, showing the distribution of participant ages and genders, as well as distribution of systolic blood pressure, diastolic blood pressure, and mean arterial pressure (MAP) across the study participants (extracted via continuous CNAP measurements).}
\label{fig_Exp_Sum}
\end{figure}

The study involved 23 healthy participants (22 to 65 years, 16~male, 7~female, cf. Fig.~\ref{fig_Exp_Sum}).
Fig.~\ref{fig_Exp_Sum} shows their baseline systolic and diastolic blood pressure measurements, demonstrating the physiological diversity of the study population.

This diversity was also present in their blood pressure readings, which ranged from 82 mmHg to 151 mmHg for systolic BP, 54 mmHg to 93 mmHg for diastolic BP, and 63 mmHg to 112 mmHg for MAP. This wide range increases the chance of observing a range of pulse wave morphologies, allowing for an evaluation of the multi-modal sensor's ability to track these different patterns accurately. This variation in participant physiology provides a basis for assessing the sensor's performance across a spectrum of physiological conditions.

Participants provided written informed consent prior to enrollment, and the study protocol received approval from the ethics committee of Aalto University (approval ID D/10330/03.04/2024).



\subsection{Experimental Setup}
We performed experiments in a room with a controlled temperature of 23 degrees Celsius.
Participants were seated comfortably in an adjustable-height chair so that their right wrist, where the multimodal sensor prototype was placed, was aligned with heart level.
A CNAP (Continuous Non-invasive Arterial Pressure) Monitor 500 served as gold standard reference for continuous blood pressure measurements.
An appropriate arm cuff was fitted to the left arm.
After selecting the correct size, finger cuffs were attached to the index and middle fingers of the left hand.
The CNAP system performed self calibration, and participant physiological information (weight, height, age, gender) was provided to the device.

The multimodal sensor prototype was carefully positioned on the right wrist over the radial artery (located by palpation to maximize signal acquisition).
The tightness of the wrist strap securing the prototype was adjusted to ensure consistent skin contact without impeding blood flow.
Data from mmWave radar and PPG sensor were acquired simultaneously at a sampling rate of 200 Hz to ensure capturing detailed arterial pulse waveform morphology.
Participants were instructed to remain still and minimize movements throughout the data collection period to minimize motion artifacts.


\subsection{Signal Processing}
\begin{figure*}
\includegraphics[width=1\textwidth]{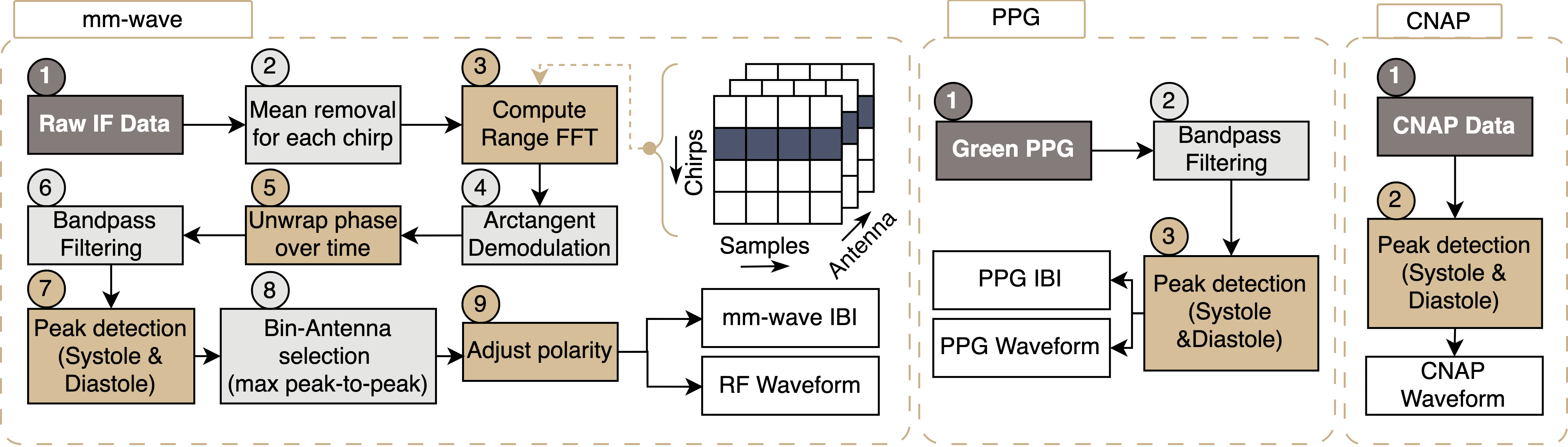}
\caption{Digital signal processing flowchart.\label{DSP_Flowchart}}
\end{figure*}

The overall signal processing flow is schematically represented in Fig.~\ref{DSP_Flowchart}.
The initial signal processing stage involved the acquisition of raw Intermediate Frequency (IF) signals, characterized by a four-dimensional data structure containing frames, antennas, chirps per frame, and samples per chirp.
Subsequently, a mean removal operation was applied across each chirp to mitigate the presence of DC bias and accentuate the subtle variations indicative of pulsatile activity. Following this, a Fast Fourier Transform (FFT) was performed along the sample dimension for each chirp to extract range-resolved information.
From the resulting frequency-domain data, the amplitude and phase components of the signal were computed.
To address inherent discontinuities arising from the arctangent function in phase estimation, a temporal phase unwrapping algorithm was implemented. Subsequently, a fourth-order Butterworth bandpass filter, characterized by lower and upper cutoff frequencies of 0.5 Hz and 8 Hz respectively, was applied to the data. This filtering step served to attenuate low-frequency baseline drift and high-frequency noise components.

Further processing involved MATLAB peak detection algorithms to the filtered phase signal to identify systolic and diastolic peaks.
To optimize signal fidelity, the peak-to-peak value of the pulsating signal was computed across all range bins and antenna elements, and the combination showing the highest peak-to-peak amplitude was selected for further analysis.
The identified diastolic peak was strategically utilized to detect potential signal inversions and to ensure the consistent polarity of the arterial pulse wave derived from the mmWave sensor.

For the analysis of the PPG data, one of the green channels of the raw signals was selected.
Subsequently, the same bandpass filter that was applied to the mmWave radar data was also implemented on the PPG signal.
The remaining steps are peak detection analysis which went through the same pipeline as mmWave signal processing described above to obtain the diastolic peaks.
Regarding the Continuous Non-invasive Arterial Pressure (CNAP) data, peak detection was applied to identify the systolic and diastolic peaks to determine blood pressure values.
The mean arterial pressure (MAP) was calculated using the formula\cite{RN19}:
\begin{equation}
    \text{MAP} = \text{DBP} + \frac{\text{SBP} - \text{DBP}}{3}
\end{equation}
For synchronization of the different sensors, the diastolic peak can serve as a temporal marking point for matching the mmWave, PPG, and CNAP. Also, the IBI were extracted based on the temporal separation between successive diastolic peaks for mmWave and PPG sensors. For CNAP, the IBI was directly reported by the CNAP system, which served as the gold standard, thus eliminating the need for estimation from the CNAP data itself. 

The signal morphology was extracted based on the waveform segments between diastolic peaks for all the sensor systems. The signal morphology derived from the PPG, mmWave radar, and the gold standard CNAP device goes through a normalization process, for both amplitude and time domains. This normalization aimed to mitigate the effects of variations in heart rate and signal amplitude, thereby enabling a meaningful comparison of the different sensing modalities.

\section{Results}

\begin{figure}
\centerline{\includegraphics[width=1\linewidth]{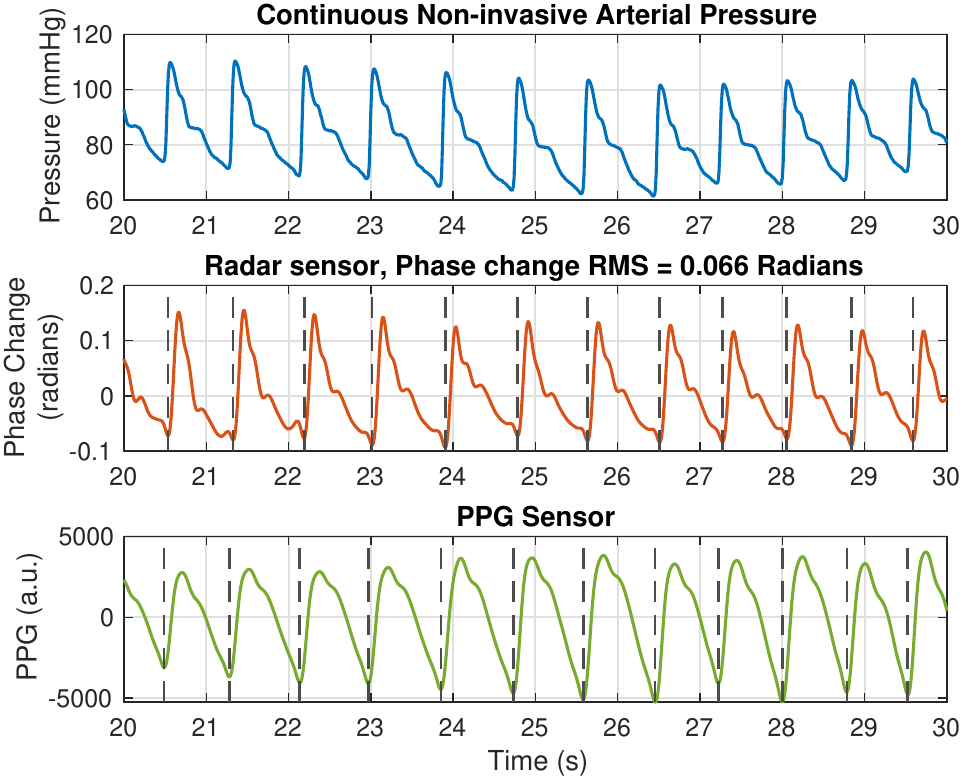}}
\caption{Example simultaneous pulse waveforms from Subject 1 over a 10-second period, showing CNAP, radar phase changes, and PPG signals, with detected diastolic peaks indicated by dashed lines.}
\label{fig_sub1_3plt}
\end{figure}

\begin{figure}
\centerline{\includegraphics[width=1\linewidth]{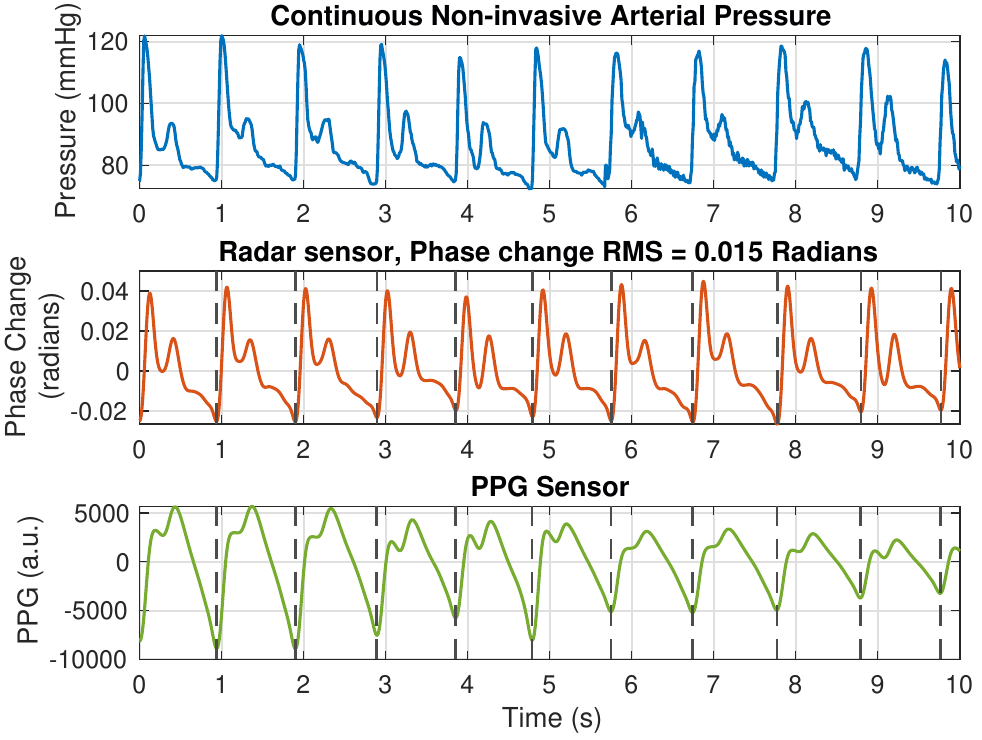}}
\caption{Example simultaneous pulse waveforms from Subject 3 over a 10-second period, showing CNAP, radar phase changes, and PPG signals, with detected diastolic peaks indicated by dashed lines.}
\label{fig_s3}
\end{figure}

Inter-subject variability in arterial pulse wave morphology obtained by mmWave radar, PPG, and the reference Continuous Non-invasive Arterial Pressure (CNAP) for two representative subjects (Subject 1 and Subject 3) over a 10-second time period is visually presented in Fig.~\ref{fig_sub1_3plt} and Fig.~\ref{fig_s3}.
In particular, the diastolic peaks are marked with dashed lines.

\begin{figure}
\centerline{\includegraphics[width=1\linewidth]{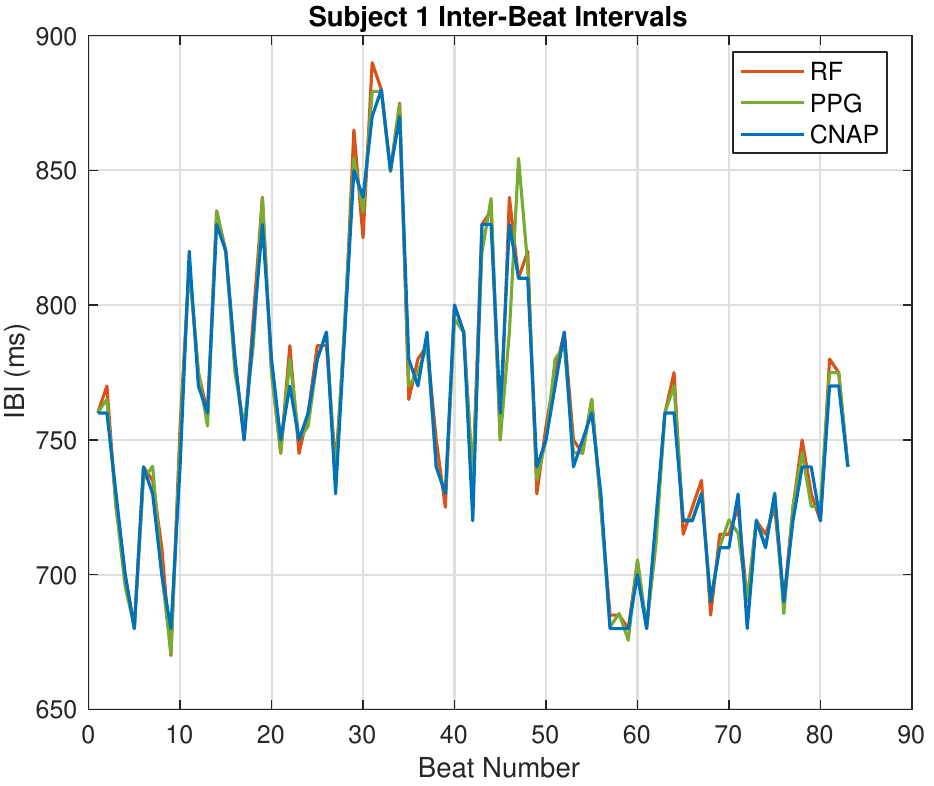}}
\caption{Inter-beat intervals (IBI) of Subject 1, showing the beat-to-beat variation in pulse timing extracted from CNAP, radar, and PPG signals.}
\label{fig_IBI}
\end{figure}

The temporal dynamics of the IBI, derived from the continuous recordings across the entire experimental duration, are illustrated as three distinct time-series in Fig.~\ref{fig_IBI} for Subject1.
These signals depict the beat-to-beat variation in the time interval between consecutive diastolic peaks as measured by the mmWave radar, the PPG sensor, and the CNAP gold standard.

\begin{figure}
\centerline{\includegraphics[width=1\linewidth]{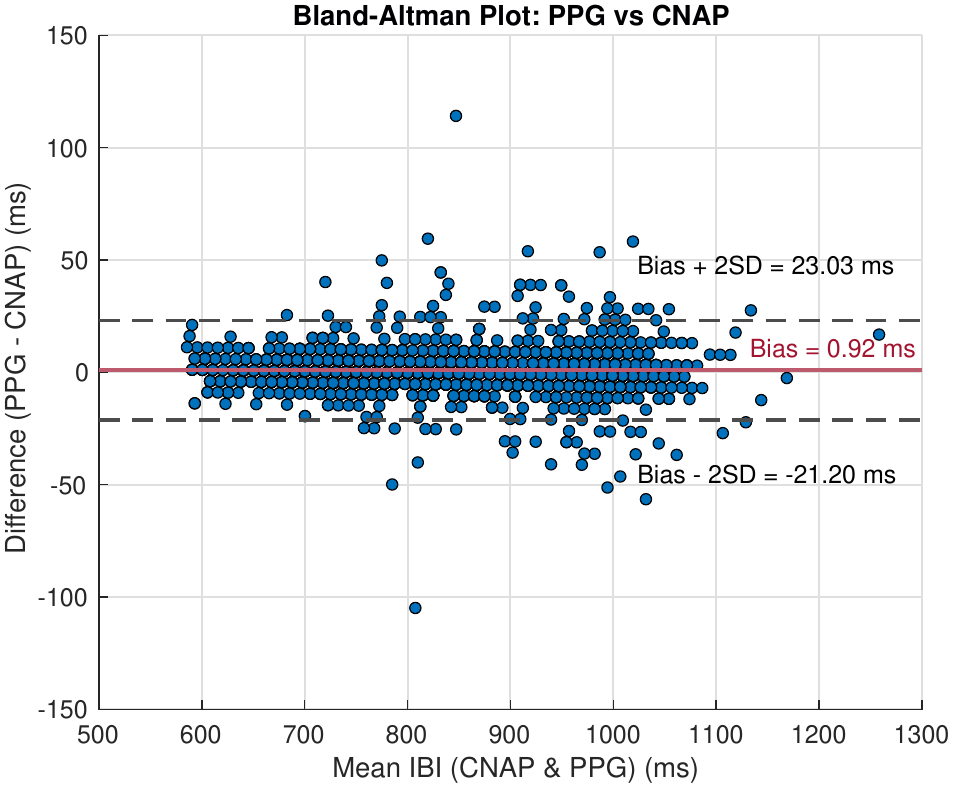}}
\caption{Bland-Altman plot comparing IBI measured by the PPG and the CNAP. The plot shows the differences between the two methods against the mean values. The mean bias and the limits of agreement (±2SD) are shown as reference lines.}
\label{fig_IBI_PPG}
\end{figure}

To quantitatively assess the agreement between the PPG-derived IBI and the CNAP reference IBI, a Bland-Altman analysis was conducted (cf. Fig.~\ref{fig_IBI_PPG}), which displays the difference between the two IBI measurements plotted against their mean value.
The 95\% limits of agreement, representing $\pm 2$ standard deviations (SD), are indicated by horizontal dashed lines, and the computed bias is represented by a red line.

\begin{figure}
\centerline{\includegraphics[width=1\linewidth]{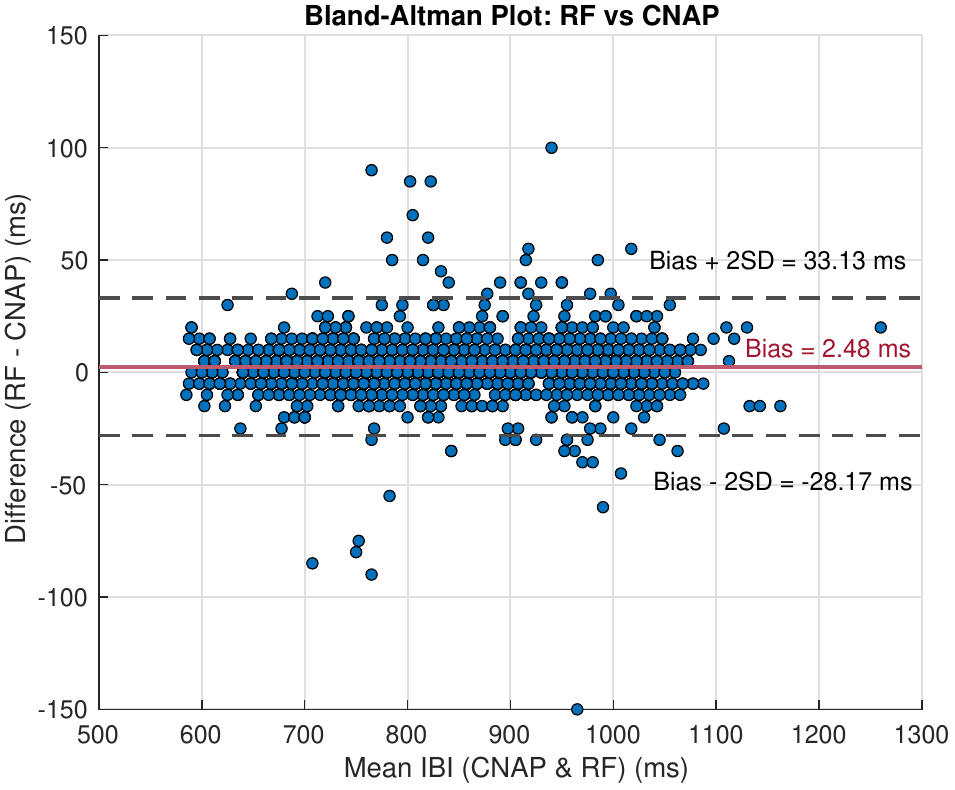}}
\caption{Bland-Altman plot comparing IBI measured by the RF and the CNAP. The plot shows the differences between the two methods against the mean values. The mean bias and the limits of agreement (±2SD) are shown as reference lines.}
\label{fig_IBI_RF}
\end{figure}

Similarly, a Bland-Altman analysis comparing the IBI obtained from the mmWave radar and the CNAP reference is presented in Fig.~\ref{fig_IBI_RF}.

The IBI, estimated for all test subjects across every heartbeat, yielded a total of 1561 data points measured by both mmWave radar and PPG.
These data points can be validated against the CNAP, providing metrics like IBI. The 95\% limits of agreement indicate that the IBI readings from the PPG sensor can be up to approximately 0.92 ms (bias) - 21.20 ms lower to 0.92 ms (bias) + 23.03 ms higher than expected.
Similarly, the IBI readings from the mmWave radar sensor can be up to roughly 2.48 ms (bias) - 28.17 ms lower to 2.48 ms (bias) + 33.13 ms higher than the expected value. The calculated bias value for PPG was 0.92 ms, and for the mmWave radar sensor, it was 2.48 ms.



\begin{figure*}
\hspace{-19mm} 
\includegraphics[width=1.1\textwidth]{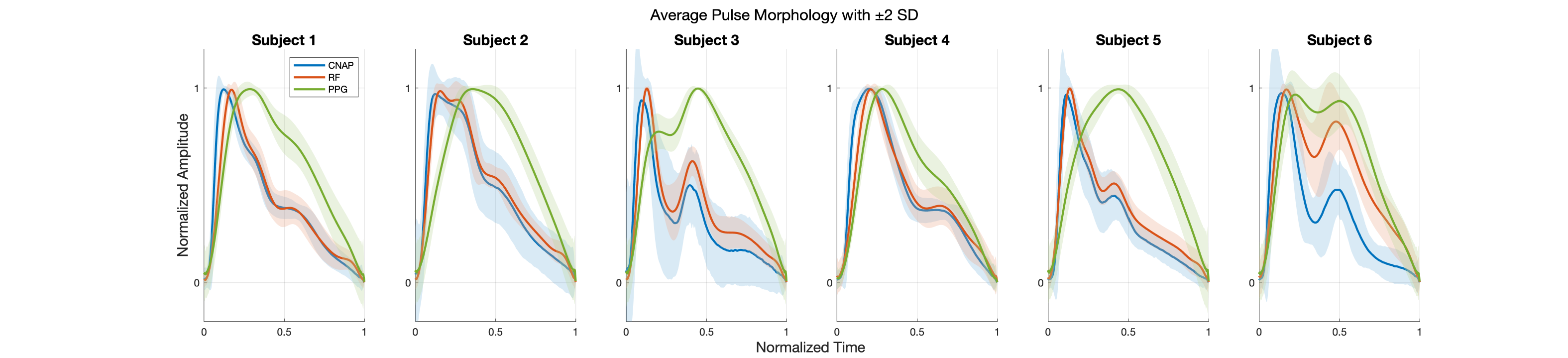}
\caption{Average normalized pulse waveforms from CNAP, RF, and PPG signals across all detected beats for different subjects. The shaded areas represent $\pm$2 standard deviations, indicating the beat-to-beat variability in pulse morphology for each modality.}
\label{fig_mrf}
\end{figure*}

A representation of the average normalized pulse waveform samples for a subset of six participants, selected based on visually inspected differences in their pulse wave morphology, is shown in Fig.~\ref{fig_mrf}. The standard deviations provide a measure of the intra-subject variability of the signals across different cardiac cycles during the experiment, offering insights into the consistency of the pulse waveform within each individual.

The quantitative analysis of arterial pulse waves from different sensor technologies was investigated and demonstrated in Table~\ref{tab:waveform_morphology}. Inflation points are the number of points where the derivative of the signal is equal to zero and averaged through the experiment and the area under curve (AUC) is the integration of the wave. Table~\ref{tab:ref_vs_ppg_rf} shows the difference between the metrics estimated by different technologies versus gold standard device (CNAP).

\begin{table*}[ht]
\centering
\renewcommand{\arraystretch}{1}
\begin{tabular}{|l|c|c|c|}
\hline
\textbf{Metric} & \textbf{PPG} & \textbf{Reference (CNAP)} & \textbf{Radar} \\
\hline
Mean Number of Inflection Points $\pm$ STD & 1.8044 $\pm$ 1.0908 & 2.8005 $\pm$ 1.4414 & 3.1183 $\pm$ 1.064 \\
\hline
Mean Area Under the Curve $\pm$ STD & 0.6165 $\pm$ 0.03774 & 0.3628 $\pm$ 0.0546 & 0.4441 $\pm$ 0.0601 \\
\hline
\end{tabular}
\caption{Comparison of waveform morphology metrics across reflectance PPG, reference arterial pulse, and radar-based waveforms.}
\label{tab:waveform_morphology}
\end{table*}

\begin{table*}[ht]
\centering
\renewcommand{\arraystretch}{1}
\begin{tabular}{|l|c|c|}
\hline
\textbf{Metric} & \textbf{Reference–PPG} & \textbf{Reference–Radar} \\
\hline
Mean Difference in Number of Inflection Points (p-value) & -0.9960 ($<0.001$) & 0.3178 ($<0.001$) \\
\hline
Mean Difference in Area Under the Curve (p-value) & 0.25365 ($<0.001$) & 0.08123 ($<0.001$) \\
\hline
Mean Cosine Similarity $\pm$ STD & 0.8572 $\pm$ 0.0459 & 0.9662 $\pm$ 0.0260 \\
\hline
\end{tabular}
\caption{Statistical comparison between reference waveforms and those derived from PPG and radar.}
\label{tab:ref_vs_ppg_rf}
\end{table*}

\section{Discussion}
It is observable that the mmWave sensor is capable of tracking the pulse wave morphology, shows a noticeable similarity to the CNAP arterial pressure, which serves as a gold standard. This implies that the signal in the phase variation of the RF sensor correlates with the pressure waveform, suggesting that by tracking the mechanical deformation of the tissue, it is possible to obtain a morphology comparable to the CNAP arterial pressure waveform. The phase variation of the radar presents a stable signal, and as visible in Fig.~\ref{fig_sub1_3plt} and ~\ref{fig_s3}, the pulse waveform remains stable throughout the heartbeats.

Upon inspecting the pressure pulse waveforms from CNAP, it is noticeable that reflections occur after the systolic peak, appearing as some peaks in the waveform. The dicrotic notch is also a point where the waveform starts to rise again during heart relaxation, continuing with a main reflection peak. These reflection peaks can be a signature in the waveform originating from the mechanical properties of the arteries. The mmWave sensor can successfully follow these reflection signals in both amplitude and time.

The green PPG signal can potentially miss the reflections due to the transition of blood from arteries to capillaries, and since the decay of the PPG pulse is slower than the other two sensor modalities, the reflection may not be clearly visible. Also, the dicrotic wave can vary in amplitude, being at a similar level to the systolic peak, lower, or even higher, which is not always consistent with the scenario observed in the standard pressure waveform.

The Bland-Altman plots for the IBI derived from both the mmWave and PPG sensor indicate an acceptable level of agreement with the CNAP reference device. The observed low bias values for both sensor systems further suggest minimal systematic error in their IBI estimations. The IBI estimation is predicated on the MATLAB peak detection algorithm, typically involving the identification of the foot of the upstroke leading to the systolic peak. However, the morphology of the mmWave radar signal may contain reflection peaks, which could potentially be misidentified by a basic peak detection algorithm as diastolic peaks, leading to uncertainties in IBI estimation. This difference in waveform morphology between PPG, which primarily reflects blood volume changes, and mmWave radar, which captures tissue deformation due to arterial pulsation, could account for the subtle variations observed in their respective Bland-Altman IBI plots. This reasoning underscores the potential necessity for the development and implementation of more advanced peak detection algorithms specifically suitable for the unique characteristics of mmWave radar pulse waveforms.

The mmWave sensor demonstrates a significant capability in capturing the details of the arterial pulse wave, including the Late Systolic Augmentation Wave and the secondary pressure increase occurring after the dicrotic notch. The presence of the augmentation wave is linked to individual vascular characteristics, and it may not always be evident in measurements obtained, even from the CNAP system, as the example participant in Fig.~\ref{fig_s3}, where this absence is noted. Notably, the mmWave sensor's phase change demonstrates a close tracking of the pulse wave morphology observed by CNAP, accurately reflecting both the systolic and dicrotic waves with correct timing. As the user mentions in relation to a hypothetical Fig.~\ref{fig_sub1_3plt}, while CNAP may show a continuation of the systolic wave with a low-amplitude augmentation and dicrotic wave, the mmWave sensor detects these key components and can potentially resolve the subtle augmentation wave across several heartbeats.
Further advancements in radar configuration and signal processing techniques can likely enhance the precision with which these minor physiological variations are consistently monitored. 
This highlights a crucial benefit of cuff-less mmWave sensing: its ability to provide continuous monitoring of the arterial pulse wave, which is a vital step towards applications such as improved sleep monitoring and the continuous assessment of cardiovascular health. The high sensitivity of mmWave radar to arterial pulsations positions it as a promising tool for in-depth cardiovascular assessment, potentially offering a more detailed insight into pulse wave morphology.

The comparison of normalized arterial pulse waveforms obtained from three sensing modalities (CNAP, mmWave radar, and PPG) in Fig.~\ref{fig_mrf} illustrates six distinct examples of pulse wave morphologies. The temporal alignment between the CNAP and mmWave waveforms is notably higher when contrasted with the PPG waveform. The PPG waveform typically presents as broader with a more gradual decay, whereas CNAP and mmWave exhibit sharper signals with more defined upslopes, more closely resembling the expected characteristics of a central arterial waveform1 . Both the CNAP and mmWave signals effectively detect the dicrotic notch, which is the secondary dip observed in the descending slope of the pulse wave.
In contrast, the PPG signal often displays a smoother decay. The variability in the signals shows the consistency and repeatability of the sensor technology. When compared to PPG, the mmWave radar demonstrates a very similar pulse wave morphology to CNAP, and the confidence intervals associated with the mmWave data suggest a stable sensor technology. 

The quantitative analysis in Table~\ref{tab:waveform_morphology} shows that the mean number of inflection points in PPG, CNAP, and radar are 1.8, 2.8, and 3.1, respectively, which shows that the mmWave sensor is capable of detecting the reflected waves similar to the reference device. The mean AUC, which was estimated based on normalized signals, demonstrated that the reflective PPG has a higher value, indicating the slow decaying of the PPG signal. The mean AUC of reference and CNAP are 0.36 and 0.44, show the similarity between these two waveforms. According to the statistical comparison in Table~\ref{tab:ref_vs_ppg_rf}, the reflective PPG is missing 1 inflection point on average compared to the reference, and the mmWave-based sensor can detect the inflection points which was detected by the reference device. The mean AUC comparison demonstrates a 0.25 higher value compared to the reference which emphasize on the slow decay of PPG, while the difference between mmWave and reference is 0.08, which shows an acceptable level of agreement for AUC. Finally, the cosine similarity for reference against PPG and mmWave is 0.85 and 0.96, shows that the mmWave-based signal is more aligned with the reference waveform.

\section{Conclusion}
We evaluated the prospective of a frequently used sensor in wearable technology, PPG, side by side with mmWave radar as a mobile and ubiquitous smart system.
The dynamic signals tracked by these sensors were tested across a diverse range of participants and validated against a cuff-based reference solution.
The study demonstrated the potential of the mmWave sensor to capture the arterial pulse wave with a morphology similar to that obtained from the cuff-based reference.
Both PPG and mmWave achieved a good degree of accuracy in tracking the IBI which can be employed for respiration rate and stress level monitoring.

PPG operates by detecting changes in blood volume, while mm-Wave radar is sensitive to the mechanical deformation of tissues resulting from arterial pulsation. This fundamental difference suggests that by combining the information obtained from both PPG and mmWave sensors, it may be possible to gain a more comprehensive understanding of cardiovascular mechanics, potentially enabling the estimation of parameters such as arterial stiffness. This interdisciplinary approaches to smart computing in wearables can provide more advanced cardiovascular health signals, providing more awareness about the health condition of the user.

\section*{LIMITATIONS AND FUTURE WORKS}
The current study's conclusion is based on 23 test participants, and even though the statistical significance of the results is acceptable, it is recommended that there be more variation in the test participants and a better balance between categories like gender and health conditions. In addition, a more in-depth analysis of the mmWave-based waveform morphology is to be conducted. Future research will establish a connection between the details present in the mmWave signal and the mechanical properties of the arteries, such as arterial stiffness. Moreover, the complementary nature of blood volume information from PPG and mechanical deformation data from mmWave holds the potential for more advanced analyses of overall cardiovascular health.


%
%

\begin{acks}
We acknowledge partial funding by the Finnish Doctoral Program Network in Artificial Intelligence, AI-DOC
(with decision number VN/3137/2024-OKM-6).
The authors gratefully acknowledge ams-osram for generously providing the optical biosensing development kit utilized in this research.
\end{acks}

\bibliographystyle{ACM-Reference-Format}
\bibliography{sample-base}



\end{document}